\documentclass[12pt]{article}
\usepackage{graphicx}

\newcommand\pubnumber{NuPhys2018-Smirnov}
\newcommand\pubdate{\today}

\def\napoli{Max-Planck-Institut f{\"u}r Kernphysik,\\
D-69117 Heidelberg, Germany}
\def\support{\footnote{smirnov@mpi-hd.mpg.de}}

\def\Title#1{\begin{center} {\Large #1 } \end{center}}
\def\Author#1{\begin{center}{ \sc #1} \end{center}}
\def\Address#1{\begin{center}{ \it #1} \end{center}}

\newcommand\pubblock{\rightline{\begin{tabular}{l} \pubnumber\\
         \pubdate  \end{tabular}}}
\newenvironment{Abstract}{\begin{quotation}  }{\end{quotation}}
\newenvironment{Presented}{\begin{quotation} \begin{center}
             PRESENTED AT\end{center}\bigskip
      \begin{center}\begin{large}}{\end{large}\end{center} \end{quotation}}
\def\Acknowledgements{\bigskip  \bigskip \begin{center} \begin{large}
             \bf ACKNOWLEDGEMENTS \end{large}\end{center}}

\def\beq{\begin{equation}}
\def\eeq#1{\label{#1}\end{equation}}
\def\eeqn{\end{equation}}

\def\beqa{\begin{eqnarray}}
\def\eeqa#1{\label{#1}\end{eqnarray}}
\def\eeqan{\end{eqnarray}}

\let\bar=\overbar

\def\Dslash{\not{\hbox{\kern-4pt $D$}}}
\def\dslash{\not{\hbox{\kern-2pt $\del$}}}

\def\msb{{\bar{\ssstyle M \kern -1pt S}}}

\begin{document}
\begin{titlepage}
\pubblock

\vfill
\Title{Neutrino Mixing 
via the Neutrino Portal}

\vfill
\Author{ Alexei Y Smirnov \support}
\Address{\napoli}

\vfill

\begin{Abstract}

Relation between the lepton and quark mixings: 
$U_{PMNS} \approx V_{CKM}^{\dagger} U_X$, where $U_X$ is the BM or 
TBM mixing matrices,   implies the quark-lepton (Grand) unification  
and existence of hidden sector with certain flavor symmetries. 
The latter couples to the visible sector  
via the neutrino portal and is responsible for $U_X$, as well as for smallness of neutrino mass.  
GUT ensures appearance of $\sim V_{CKM}$ in the lepton mixing. 
General features 
of this scenario (inverse or double seesaw, screening of the Dirac structures, basis fixing symmetry)  
are described and two realizations 
are presented.  The high energy realization is based on $SO(10)$ GUT with the hidden 
sector at the Planck scale. The low energy realization includes  
the 100 TeV  scale $L-R$ symmetry and the hidden sector at the keV - MeV scale.  
\end{Abstract}

\vfill

\begin{Presented}
NuPhys2018, Prospects in Neutrino Physics, \\

Cavendish Conference Centre,\\
London, UK, December 19 - 21, 2018
\end{Presented}

\vfill

\end{titlepage}

\def\thefootnote{\fnsymbol{footnote}}

\setcounter{footnote}{0}

\section{Introduction}


It seems, another phase in developments of the field   
related  to the discovery of large lepton mixing  and 
wide exploration of the non-abelian discrete flavor symmetries 
is nearly over. No simple and convincing model has been 
proposed which could explain small quark mixing and 
large lepton mixing in the same framework.  
Several different symmetries with {\it ad hoc} charge prescriptions, 
non-renormalizable interactions, 
complicated flavons content, {\it etc.} are some generic features  of the models. 
The bottom line is that in understanding of neutrino mass and mixing 
we are not far from the very beginning, that is, from 
experimental results. 

In this connection the guidelines could be that  

1. After all, the Grand Unification is the best proposed physics beyond the Standard Model (SM).   
It provides unification of forces: explanation of  
why the strong interactions are strong, the weak interactions are weak, and the EM interactions are as they are.  
GUT unifies quarks and leptons and gives explanation of the SM symmetry charges.  
SO(10)  perfectly embeds all known fermions including RH neutrinos
in a single 16-plet. 

The simplest versions of GUT's predicted 
\begin{equation}
m_q \sim m_l \sim  m^{\nu}_D ~~{\rm or} ~~ m_l \sim m_d, ~~ m^\nu_D \sim m_u.   
\label{eq:mass}
\end{equation}
Beauty of the seesaw mechanism is that it allows to reconcile
relations (\ref{eq:mass}), {\it i.e.} ``normal'' values of the 
Dirac Yukawa couplings of neutrinos, and  smallness 
of neutrino mass with  only one assumption --  existence of large Majorana 
masses of the RH neutrinos, $M_R$. Furthermore, $M_R \sim M_{GUT}$.  
In many models the ``hybrid'' seesaw  is employed which uses two assumptions
(and in this sense less attractive): 
large Majorana masses of the RH neutrinos and 
smallness  of the Dirac Yukawa couplings (Dirac masses). 

2. Existing GUT picture is not complete 
(hierarchy problem, proton decay, {\it etc.}),  
something important is still missing, but adding these ``extra" may produce small 
perturbation of the main picture for visible sector.
E.g. hidden sector interacting via different portals may exist, which is 
also needed for explanation of dark matter, inflation, {\it etc.} 
 
3. Testability, especially in forthcoming and planned experiments,  
is not the problem of Nature. It is our problem.
Simplicity, minimality,  symmetry still have great value.

\section{Scenario: $\nu-$mixing from the hidden sector}

Starting point  is that the data are in a good agreement with the relation 
\cite{Giunti:2002ye,raidal,Minakata:2004xt,xing}:
\begin{equation}
U_{PMNS} = U_l^{\dagger} U_X, 
\label{eq:relation}
\end{equation}
where $U_l \approx V_{CKM}$ is the quark mixing matrix  
and 
$U_X \approx U_{BM}$ or $U_{TBM}$ are the BM- or TBM- mixing matrices.   
The diagonal matrix of phases $\Gamma_\alpha$ can be attached to $U_X$.  

The equality (\ref{eq:relation}) leads to prediction $\sin^2 \theta_{13} = 0.5 \sin^2 \theta_C$,  
and in general, for  
$U_{X} = \Gamma_\alpha U_{23}(\theta_{23}^X) U_{12}(\theta_{12}^X)$,  
one can obtain relation between the observables \cite{Minakata:2004xt}:
\begin{equation}
\sin^2 \theta_{13} = \sin^2 \theta_{23} \sin^2 \theta_C [1 - O(\sin^2 \theta_C)]. 
\label{eq:13pred}
\end{equation}
The present experimental status of this relation  is summarized in 
Fig.~\ref{fig:fit}.   
The relation can be modified due to RGE running  if it is  
fixed at high (GUT) scale.  Further $\sim  20\%$ 
correction to $\sin^2 \theta_{13}$ 
can be due to deviation of  $\theta_{12}^l$ from $\theta_C$:  $\theta_C \rightarrow \theta_{12}^l$ 
in (\ref{eq:13pred}). 
This brings the prediction to the best fit point. 

\begin{figure}[htb]
\centering
\includegraphics[height=2.0in]{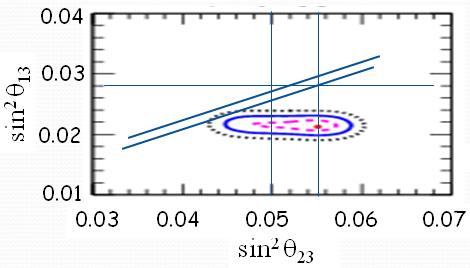}
\caption{Relation between the 1-3 and 2-3 leptonic mixings according to Eq. (\ref{eq:13pred})  
for $\theta_{12}^l = \theta_C$. Two lines show the band of predictions obtained by varying the phases 
in $\Gamma_\alpha$. 
The $1\sigma$, $2\sigma$ and  $3\sigma$ allowed regions are taken from the global fit \cite{capozzi}.}
\label{fig:fit}
\end{figure}

The form of equality (\ref{eq:relation})  implies  
that two different contributions from two different sectors 
of theory with different  symmetries are involved  
in generation of the lepton mixing:

$V_{CKM}$ follows from common sector for quarks and leptons, which gives  
Eq. (\ref{eq:mass}).  
This requires the  $q - l$ unification, GUT.  
The CKM physics is characterized by hierarchy of masses and mixings 
as well as relations between masses  and mixing which can be achieved  with,  
{\it e.g.}, Froggatt-Nielsen mechanism.

$U_X$ originates from new sector related to neutrinos   
via what we call now the neutrino ``portal''. 
It is  responsible for large (maximal) neutrino mixing and smallness 
of neutrino mass. It can have special symmetry 
which leads to the  BM or TBM mixing. \\

\begin{figure}[htb]
\centering
\includegraphics[height=2.1in]{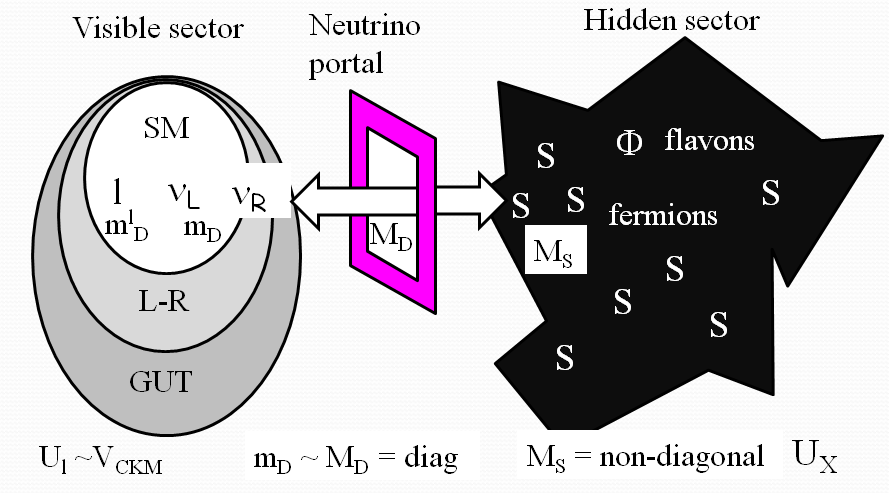}
\caption{Neutrino mixing via the neutrino portal. Shown is the general setup. 
The lepton mixing has two different sources.}
\label{fig:portal}
\end{figure}

General  setup of this scenario  (Fig. \ref{fig:portal}) is the following.  

\begin{itemize}

\item
Visible sector  contains particles of the Standard Model:  $l$, $\nu_L$, 
as well as $\nu_R$ with mass matrices $m_l$, $m_D^\nu$. 
It can be embedded into the $L-R$ symmetry model 
and then GUT. This sector produces $V_{CKM}$. 

\item
Neutrino portal: $\nu_R$ and singlet fermions 
$S$ have Dirac mass terms which form  the matrix $M_D$. 

\item 
Hidden sector: 
Apart from $S$ it contains flavons - scalar 
fields with non-zero flavor charges  which couple with $S$. Flavons  develop non-zero VEV's,  
break the flavor symmetry, and  generate non-diagonal mass matrix $M_S$ which   
is the origin of (diagonalized by) $U_X$. 

Concerning scale of the hidden sector, $M_S$,  
there are two extreme possibilities: high scale, $M_S \sim M_{Pl}$,  
or low scale, $M_S \sim \mu =$ (keV - MeV).

\end{itemize} 

In this scenario all the interactions are renormalizable and it is easy to realize 
flavor symmetries\footnote{
Notice that  more economic version without singlets 
and with usual seesaw type-I also 
has  structure which does not exist in the 
quark sector and so can be responsible for the 
difference of mixing. This, however,  
does not allow to disentangle the CKM and 
$X-$ sectors and therefore implement symmetries.}.

In general, there can be  many singlets in the hidden sector  
but some of them decouple 
or their effects can be reduced to the effect of three singlets. 
For three  $S$ which couple to neutrinos  
the mass matrix is 
\begin{equation}
{\cal M} =
\left(
\begin{array}{ccc}
0  & m_D^T  & {m_D'^T}\,  \\
 m_D  & 0  & M_D  \\
m_D'  & M_D^T & M_S
\end{array} 
\right)\, . 
\label{eq:neutral_mass_1}
\end{equation}
Block diagonalization of ${\cal M}$ gives the mass matrix of light neutrinos  
\begin{equation}
m_\nu  =  m_D^T  M_D^{-1T} M_S M_D^{-1} m_D - 
(m_D^T M_D^{-1T} m_D' + m_D'^T  M_D^{-1} m_D). 
\label{eq:doublinear}
\end{equation}
We consider the situations when 
the double or inverse seesaw \cite{Mohapatra:1986bd} (the first term in (\ref{eq:doublinear})) 
dominates,  while  the linear seesaw (the second term) is suppressed. 

If $m_D = A M_D$, where $A$ is a constant,  
the first term  of Eq. (\ref{eq:doublinear}) gives 
$$
m_\nu = A^2 M_S, 
$$ 
In this case structure of $m_\nu$  
is  determined by  $M_S$, and it does not depend 
on structure the Dirac mass matrices (what was called in \cite{screening} 
the Dirac screening). 
The screening allows to disentangle the sectors, and at the same time -  transfer 
the flavor information from the hidden sector to the visible one.

In general, the information about mixing in the hidden sector should be communicated  
to the visible one. For this, the simplest possibility 
is to introduce symmetry which fixes bases in all three sectors,  
and the simplest basis fixing symmetry is $G_{basis} = Z_2 \times Z_2$ \cite{Ludl:2015tha}. 
Indeed, the $G_{basis}$ transformations  of the fermionic multiplets and singlets  
$(-, -), (+, -), (-, +)$ allow to  distinguish three generations. 
If the Higgs multiplets of visible sector are singlets of $G_{basis}$,  then 
$m_D \sim M_D = {\rm diagonal}$.  
To ensure the proportionality of the diagonal elements $m_D^{diag} \sim M_D^{diag}$, 
required by complete screening, one needs to introduce additional (e.g., permutation) symmetry 
or rely on further unification of $S$ and other fermions. 
Flavons $F$ are charged with respect to $G_{basis}$   
and spontaneously break $G_{basis}$, which leads to 
non-diagonal  $M_S$,  and consequently, to  
mixing $U_X$. 

$G_{basis}$ is a part of the intrinsic symmetry $(Z_2)^3$
of Majorana mass matrix  which is always present,  {\it i.e.}, 
$G_{basis}$ is given ``for free'' \cite{bajc}.
 
$M_S$ diagonalized by $U_X$ has another unbroken
 symmetry $(Z_2 \times Z_2)_H$. Thus, $U_X$ connects bases determined 
by $(Z_2 \times Z_2)_V$  and $(Z_2 \times Z_2)_H$.   
To fix $U_X$ one can assume embedding  of 
$(Z_2 \times Z_2)_V$  and $(Z_2 \times Z_2)_H$ into 
a finite discrete group (residual symmetry approach):
Using the symmetry group condition \cite{Hernandez:2012ra} one finds that 
embedding of two Klein groups leads to general expression
for elements of mixing matrix \cite{bajc}, \cite{xu}  
\begin{equation}
|(U_X)_{ij}|^2 = \cos^2 \pi \frac{n_{ij}}{p_{ij}}, ~~~~ p , n -{\rm integer}. 
\end{equation}
This expression and  the unitarity condition 
$$
\sum_i \cos^2  \pi \frac{n_{ij}}{p_{ij}} = 1, 
$$     
(and similar equalities hold for the sum over $j$) 
allow to reconstruct the matrix $U_X$ up to discrete number 
of possibilities. 
Taking into  account that elements of $U_X$  
are in general complex,  5 matrices have been found \cite{xu}. Among them are 
$U_{q/p}$,   $U_{BM}$: 
\begin{equation}
U_{q/p} = \left(\begin{array}{ccc}
1 & 0 & 0\\
0 & \cos\frac{q}{p}\pi & \sin\frac{q}{p}\pi\\
0 & -\sin\frac{q}{p}\pi & \cos\frac{q}{p}\pi
\end{array}\right), ~~~~
U_{BM} = 
\left(\begin{array}{ccc}
1 \left/\sqrt{2}\right. & 1\left/\sqrt{2}\right. & 0\\
-1/2 & 1/2 & -1\left/\sqrt{2}\right.\\
-1/2 & 1/2 & 1\left/\sqrt{2}\right.
\end{array}\right)~. 
\end{equation} 
as well as  the golden ratio matrix $U_{GR}$ which 
could be  important for phenomenology. 
In the case of $U_{BM}$ the covering symmetry is $S_4$.  $U_{TBM}$ can not be obtained.  
One could consider more complicated  basis fixing symmetry which  has different embeddings
and covering groups.

\section{High energy GUT-Planck realization}

In this realization  $M_S \sim  M_{Pl}$, we assume SO(10) GUT,  
and  the portal scale $M_D \sim  M_{GUT}$ \cite{xu}. 
The linear seesaw contribution is ``automatically" suppressed and neutrino masses 
are generated via the double seesaw.   
The RH neutrinos get mass via see-saw
$M_R  \approx M_D^T M_S^{-1} M_D$. 

%
\begin{figure}[htb]
\centering
\includegraphics[height=2.0in]{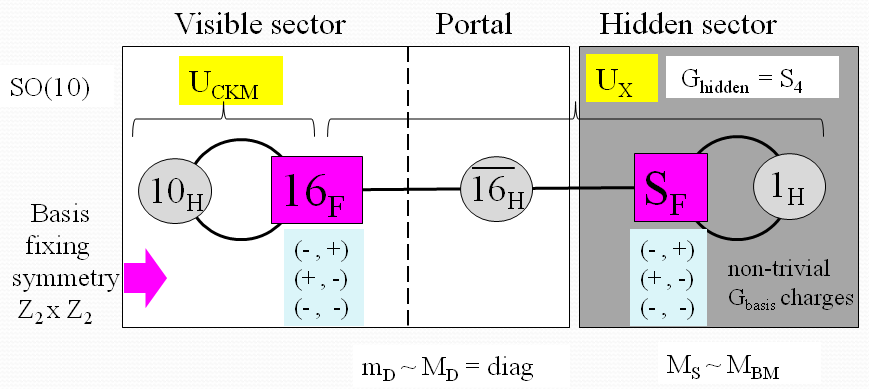}
\caption{Scheme of generation of mixing in the high scale scenario.}
\label{fig:highreal}
\end{figure}

The scheme with $G_f = G_{hidden} = S_4$ is shown in Fig. \ref{fig:highreal}. 
$G_f = S_4$ is explicitly broken down to $G_{basis} = (Z_2 \times Z_2)_V$ 
in the visible and portal sectors,  and
spontaneously - down to $(Z_2 \times Z_2)_H$ in the hidden sector.
The explicit breaking gives very small corrections to the PMNS-mixing. 


With respect to  $S_4$ the fermionic 16-plets and singlets $S$ 
transform  as $ \sim \bf 3$,  
flavon form triplet $\phi$  $ \sim {\bf 3'}$  and  doublet  
$\xi$  $ \sim {\bf 2}$,  and  other fields are  $S_4$  
singlets. The VEV alignment
$$
\langle \phi \rangle^T \sim (0, 0, 1), ~~~   
\langle \xi \rangle^T \sim  (0, 1)
$$
gives $M_S = M_{BM}$,  and consequently, $U_X  = U_{BM}$.\\


$U_l$ and $V_{CKM}$ follow from the down components of the fermion 
EW doublets. 
The required equality
$$
U_l \approx U_d = V_{CKM}
$$
should be reconciled with difference of masses at the GUT scale:
$$
m_\mu \sim 3m_s, ~~~   m_d \gg m_e. 
$$
Here, general idea is that the mass matrices have two different contributions:  
\begin{equation}
M_d = M_d^{(10)} + M_C, ~~~~   M_l = M_d^{(10)} - 3M_C, 
\label{eq:twocont}
\end{equation}
but only one contribution  dominates in generation of a given
2 fermion mixing  1-2 and 1-3.  
In (\ref{eq:twocont}) $M_d^{(10)}$ is the  contribution from $10_H$. 
It is diagonal and 
strongly hierarchical, as up quark masses 
$M_d^{(10)}  =  v_d /v_u M_u^{(10)}$, 
and therefore dominates in generation of mass and mixing of the third generation states.  

$M_C  = {\cal O}(m_s)$ is sub-dominant for the third generation but is less hierarchical 
and  therefore  dominant for the 2nd and 1st  generations.  
The matrix $M_C$ is off-diagonal, it breaks
 $G_{basis} = G_{visible}$ and produces the  CKM mixing.  The required form is 
\begin{equation}
M_{C} \approx 
\left(\begin{array}{ccc}
d_1 & f & f'\\
f & d & d'\\
f' & d' & d
\end{array}\right) ~~~~\frac{f}{d} = \sin \theta_C,   ~~ d_1 \ll d \sim d', ~~~ f \sim f'. 
\label{eq:gut-41}
\end{equation}
Interestingly,  $M_C \propto M_{BM}$,  and therefore both 
could originate from  the same Planck scale physics.
The largest elements of $M_C$ are of the order  $d, d' \sim  0.1 v_{EW} M_{GUT}/M_{Pl}$.  
$M_C$ can be generated by additional 126-plet of Higgses  with Planck
 scale mass or by effective 126: $M_{Pl}^{-1} 16_F 16_F 16_H 16_H'$
with $\langle 16_H 16_H'\rangle \sim v_{EW} M_{GUT}$ \cite{xu}. 

Fit of  the observed masses and mixing of quarks with 
(\ref{eq:twocont}) and (\ref{eq:gut-41})
gives the angles in $U_l$:
$\theta_{12}^l \sim  \theta_C$,    
$\theta_{23}^l \sim  (4 -5)^{\circ}$,    
$\theta_{13}^l \approx 1^{\circ}$. 
Using uncertainties in the quark masses and phases involved we have
$$
\theta_{12}^l / \theta_C  =  0.87 - 1.35. 
$$
According to $U_{PMNS}  =  U_l^{\dagger} U_{BM}$,  
the parameters of the PMNS matrix 
equal 
\begin{equation}
s_{13}^2 = \frac{s_{l}^2}{2},~~~
s_{12}^2 = \frac{1}{2}-\frac{\sqrt{2}c_{l}s_{l}\cos\phi_{l}}{2-s_{l}^{2}},~~~
s_{23}^2 = \frac{c_l^2}{2-s_{l}^{2}} \approx
\frac{1}{2}\left( 1 - \frac{1}{2} s_l^2\right), 
\label{eq:predict123}
\end{equation}
\begin{equation}
\sin\delta_{{\rm CP}}=-\sin\phi_{l}\left(1 + s_{l}^{2} \cos^{2}\phi_{l}\right) +
{\cal O}(s_{l}^{3}), 
\label{eq:gut-118}
\end{equation}
where $s_l \equiv \sin \theta_{12}^l$, and $\phi_l$ is the phase of the 1-2 element: 
$(U_l)_{12} = s_l e^{i \phi_l}$. 


\begin{figure}[htb]
\centering
\includegraphics[height=1.7in]{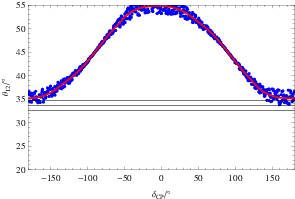} \hskip 0.2cm
\includegraphics[height=1.6in]{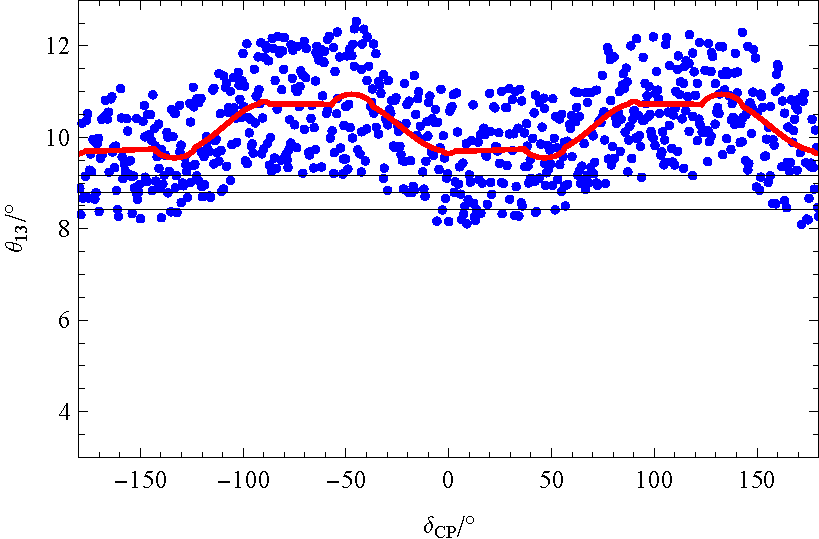}
\caption{Dependence of  12 ({\it left}) and 13 ({\it right}) mixing angles on the  CP phase.  
Blue points correspond to values of  the charged fermion masses
randomly generated within $1\sigma$ allowed regions. From \cite{xu}. }
\label{fig:1213pred}
\end{figure}

Excluding $s_l$ and $\phi_l$ from 
Eqs. (\ref{eq:predict123}) and (\ref{eq:gut-118}) we find  
relations between observables: 
$$
s_{12}^2 \approx \frac{1}{2}+ \frac{s_{13}\cos\delta_{{\rm CP}}}{c_{13}^{2}}, ~~~
s_{13} = \frac{3}{\sqrt{2}}\sin \theta_C \left|\frac{m_s - m_d e^{i\phi_d}}{m_\mu + m_e e^{i\phi_e}}\right| .
$$
where $\phi_d = \phi_d(\delta_{CP})$, $\phi_e = \phi_e(\delta_{CP})$ are known  functions of 
$\delta_{CP}$ \cite{xu}, see Fig. \ref{fig:1213pred}. 
From the figure we obtain  $\delta_{CP}  = (0.80 - 1.16) \pi$
which touches  the $1\sigma$ region from the global fit:
$(1.17 - 1.53) \pi$.  
Notice that $\cos \delta_{CP}  \approx -1$  is a generic prediction
for the BM mixing case~\cite{Petcov:2017ggy}.  
RGE can change this result,  so that $\delta_{CP}  = - 0.5 \pi$  
becomes possible. 


Tests and problems: in this realization one expects that 
(i)  flavons, new fermions and new Higgses are at the  GUT – Planck scale;  
(ii)  nothing should be observed at LHC which 
is responsible for neutrino mass generation;  
(iii) proton decays; 
(iv) new elements of theory related to the CKM physics may show up; 
(v)  the RH neutrinos have very strong hierarchy of masses;  
leptogenesis with second RH neutrino is possible \cite{DiBari:2015svd};  
(vi) other particles from the hidden (Dark) sector can be found 
such as  sterile neutrinos, DM particles, {\it  etc.}


\section{Low scale realization with the L-R symmetry}

\begin{figure}[htb]
\centering
\includegraphics[height=1.8in]{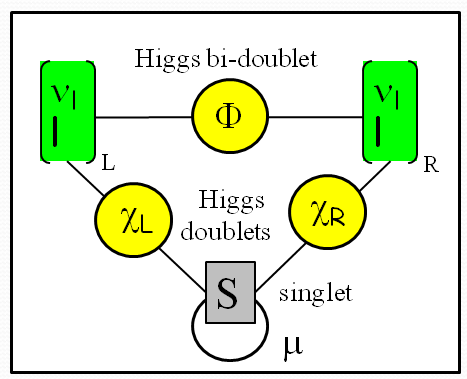} \hskip 0.5cm
\includegraphics[height=1.9in]{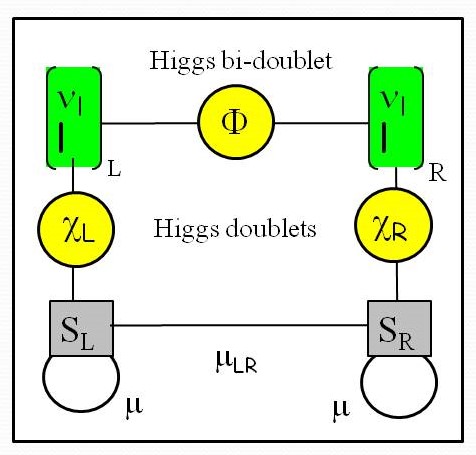}
\caption{Generation of masses and mixing in the low scale scenarios. 
{\it Left panel:} scheme with one singlet $S$ per generation. 
{\it Right panel:} scheme with two singlets, $S_L$ and $S_R$,  per generation. 
}
\label{fig:schemes}
\end{figure}

The low scale scheme with  $SU(2)_L \times  SU(2)_R  \times  U(1)_{B-L}  \times P$ symmetry 
and  one  singlet $S$  per generation \cite{brdar} is shown in  Fig. \ref{fig:schemes} {\it left}. 
Here $P$ is the parity.  
The  $B-L$ charges of the fields 
$(L_L, ~L_R,~ \chi_L,~ \chi_R, S)$ equal $(-1,~ -1,~ 1, ~ 1, ~0)$. 
For Dirac mass  matrices of Eq.(\ref{eq:mass})  small neutrino mass can be obtained using 
the inverse seesaw (ISS) mechanism  (\ref{eq:neutral_mass_1})  with $M_{S} = \mu \sim 10$ keV \cite{brdar}.  

\begin{figure}[htb]
\centering
\includegraphics[height=1.8in]{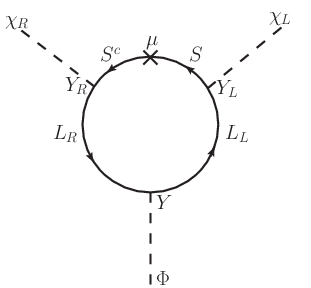} 
\includegraphics[height=1.1in]{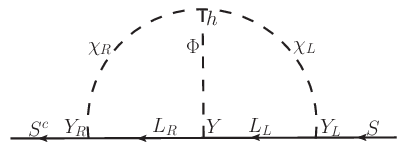}
\caption{{\it Left panel:} one loop diagram that generated $\chi_L^{\dagger} \Phi \chi_R$ interactions.
{\it Right panel:} the leading radiative correction to the Majorana mass $\mu$. From \cite{brdar}.}
\label{fig:diagrams}
\end{figure}

In general, in the low scale case,  the linear seesaw dominates. 
Due to $P$ symmetry the Yukawa couplings of two Higgs doublets are the same (in the lowest order), 
and consequently $m_D' \propto M_D$. Therefore the linear seesaw contribution reduces to   
\begin{equation}
m_{\nu}^{LSS} = 
\frac{\langle \chi_L \rangle}{\langle \chi_R \rangle}(m_D^T  +  m_D).
\label{eq:lsss}
\end{equation}
It has wrong (too strong) mass hierarchy 
and therefore should be suppressed,  which requires   
$\langle \chi_L \rangle / \langle \chi_R \rangle < 10^{-12}$. 
For this the  interactions  $h \chi_L \chi_R \phi$,   
which leads to VEV of $\chi_L$,  should have small coupling  $h < 40$ keV. 
Even  if  $h = 0$ and therefore  $\langle \chi_L \rangle = 0$ at tree level due to certain symmetry,  
the interaction term  is generated  at 1 loop (Fig. \ref{fig:diagrams} {\it left}).  
The corresponding induced VEV equals  $\langle \chi_L \rangle \sim 
 1/16 \pi^2 v_R (v_L \mu/v_R^2) $, which  satisfies the bound. Here $v_R \equiv \langle \chi_R \rangle$ and 
$v_L \equiv \langle \Phi \rangle$ is the bi-doublet VEV. 

The neutrino mass determines via the ISS the $L-R$ symmetry breaking scale. For 
$m_D^{\nu} \sim m_{top}$ it equals 
$$
v_R = \langle \chi_R \rangle  = 3.5 \cdot 10^5 ~{\rm GeV}
 \left(\frac{\mu}{0.1~ {\rm MeV}}\right)^{1/2}
\left(\frac{0.05 {\rm eV}}{m_{\nu3}}\right)^{1/2}.  
$$

As in the high scale scheme,   
at  $m_D \propto  M_D$  the Dirac structures are screened, and 
$$
m_\nu =  \xi^2 \mu ,  ~~~~ \xi \equiv \frac{m_D}{M_D} = \frac{v_L}{v_R}. 
$$
Symmetries in the $S$-sector can lead to special 
form of $\mu$, and  consequently, to  special mixing
from the hidden sector. 
The symmetry is broken (explicitly) in the portal,  
by $M_D$. But  in spite of the fact that $\mu \ll M_D$, 
corrections due to symmetry breaking (see Fig. \ref{fig:diagrams}) are small:  
$$
\Delta \mu = \frac{1}{16 \pi^2} h Y_L Y_R Y \sim 10 ~{\rm eV}, 
$$
while $\mu \gg 10$ eV for $h \sim 0.1$ MeV.


The components $N_i$  and $S_i$ form pairs  of the pseudo-Dirac leptons with  
masses and mass splittings 
$$
|M_i| \approx M_{Di} (1 + \xi^2)^{1/2},  ~~~~ \Delta M_i  = m_{ii}.
$$
Their production and decay proceed, mainly,  
via mixing in the light flavor neutrinos:
$$
\nu_f = U_{PMNS}\nu -\frac{1}{\sqrt{2}}\xi U_l^{\dagger} (N^{-} - N^{+}), 
$$
where $N^{-}$ and $N^{+}$  are the mass eigenstates. 
Thus, mixing of the heavy lepton in the flavor state 
$\nu_\alpha$ equals
\begin{equation}
|U^N_{\alpha i}|^2  = \frac{1}{2}  
\left(\frac{m_{Di}}{M_{Di}}\right)^2 |U^l_{\alpha i}|^2.
\label{eq:hlept}
\end{equation}
The dependencies (\ref{eq:hlept}) together with experimental bounds on mixing parameters of $N_i$ \cite{brdar} 
are shown in Fig. \ref{fig:fit}.  
From this figure for $m_u  \approx 2$ MeV 
we obtain  the lower bound  $M_1  > 2$ GeV,  so that   $\xi = m_u/M_1  < 10^{-3}$.  
Consequently, $M_2  > 600$ GeV, and   $M_3  > 2.5 \cdot 10^5$ GeV. 
SHiP \cite{Alekhin:2015byh} can further  improve the bound  on $M_1$ or discover $N_1$. 
Presently, there is no direct experimental bounds on $M_2$  and $M_3$. 
In future,  100 TeV collider may be sensitive to them. 
 %
\begin{figure}[htb]
\centering
\includegraphics[height=2.0in]{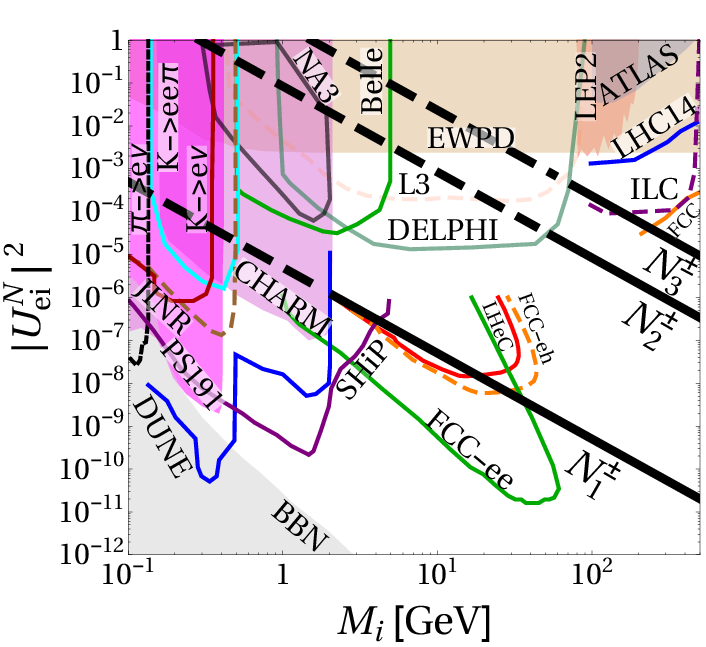}
\includegraphics[height=2.0in]{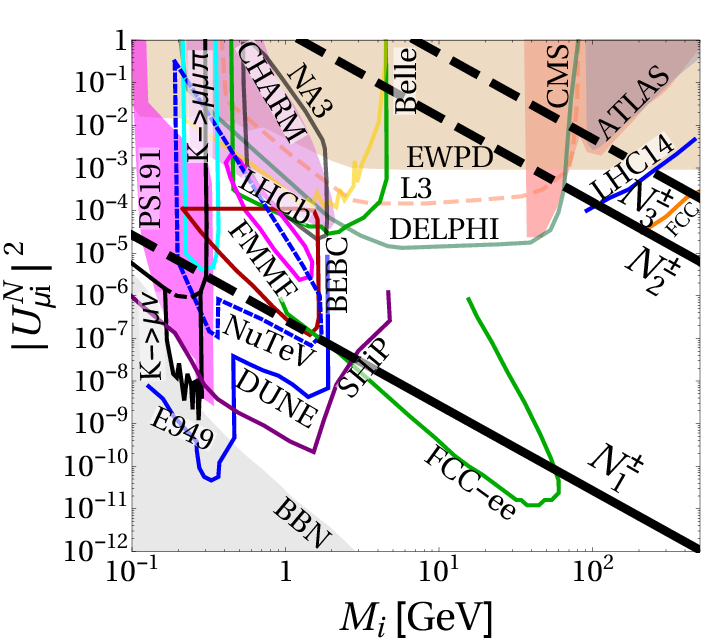}
\caption{Mixing of the heavy leptons in the neutrino flavor states $\nu_e$ ({\it left}) 
and $\nu_\mu$ ({\it right}) as functions of their masses.
Solid black lines show predictions. Colored regions and lines show bounds from the existing 
and future experiments. From \cite{brdar}.  
}
\label{fig:fit}
\end{figure}

Interesting variation of this  
scenario  is a scheme with two singlets: left and right per generation 
(see Fig. \ref{fig:schemes} {\it right}).  
It is invariant  under global $U(1)_L$ 
with  charge prescription $(L_L,~ L_R,~ S_L,~ S_R)  = (1, 1, -1, -1)$.  
This symmetry  is broken in the hidden sector by the $\mu$-terms.

Mass matrix of neutral leptons in the basis $(\nu_L, ~ N_L, ~ S_L, ~ S_R^c)$ reads 
\begin{equation}
{\mathcal M} =
\left(
\begin{array}{cccc}
  0 & m_D & m_D' & 0  \\
  m_D & 0 & 0  &  M_D     \\
  m_D' & 0 & \mu  &  \mu_{LR}  \\
   0 & M_D & \mu_{LR}^T  &  \mu
\end{array}
\right). 
\label{eq:connecting_states}
\end{equation}
Pairs of the pseudo-Dirac heavy leptons formed by $N$ 
and $S_R^C$  have similar phenomenology as before. 
After decoupling of  these heavy states  the mass matrix in the basis
$(\nu_L, S_L)$ becomes
\begin{equation}
\left(
\begin{array}{cc}
  \mu\, s^2_\xi & c_\xi\, m_D' - s_\xi \,\mu_{LR} \\
  c_\xi \, m_D' -  s_\xi\,\mu_{LR}  & \mu
\end{array}
\right).
\label{eq:2states}
\end{equation}
In contrast to the high scale scheme now light Majorana leptons  
with  masses $(10 - 100)$ keV exist which   
nearly coincide with $S_{Li}$.    
They mix very weakly with usual active neutrinos:  
$$
\sin \theta_s \approx  - \xi \frac{\mu_{LR}}{\mu}.
$$ 
If $\mu_{LR} /\mu  < 10^{-2}$, $S_L$ with mass $\sim$ 10 keV 
can  be the Dark matter candidate \cite{brdar}.


\section{Conclusion}


1. If not accidental, the relation between
the lepton and quark mixings $U_{PMNS}  = V_{CKM}^+ U_X$,  
where $U_X \sim U_{BM}$ or $U_{TBM}$  
implies Grand Unification  and existence of the hidden
sector which  has certain symmetry and  
interacts with the visible sector  via the neutrino portal.
 
2. The hidden  sector with non-abelian flavor
symmetries generates 
large neutrino mixing
of special type and is responsible for smallness
of neutrino mass.

3. The key elements of this scenario are (i) 
existence of two sectors  with different symmetries;  
(ii) the basis fixing symmetry which 
communicates flavor information from the hidden sector to the visible one. 

4. The high scale realization of such a
scenario  is the $SO(10)$  GUT  with hidden 
sector at the Planck scale.  
Neutrino masses are generated by  the double seesaw. The residual symmetry approach
can lead to the BM –mixing for $U_X$.

5. Similar scenario  can be realized  at low energies in   
the scheme with  $L - R$ symmetry and the inverse seesaw.
The scale of  $L - R$ symmetry  breaking is about 300 TeV. 
The pseudoDirac heavy leptons can be searched at
existing and future accelerator experiments. 
In version with two singlets per generation the kev mass scale 
leptons exist which can be candidates for Dark Matter particles. 


%



\Acknowledgements
 
I am grateful to Xun-Jie Xu and Vedran Brdar 
for collaboration.


\end{document}